\documentclass[11pt]{iopart}

\usepackage{mathrsfs}
\usepackage{iopams}
\usepackage{cite}

\def\udot{\dot{u}}
\def\3nab{\tilde{\nabla}}
\def\lgl{\langle}
\def\rgl{\rangle}
\def\la {\langle}
\def\ra {\rangle}
\def\c{\mbox{\,curl\,}}
\def\be {\begin{equation}}
\def\ee {\end{equation}}
\def\bea {\begin{eqnarray}}
\def\eea {\end{eqnarray}}

\newcommand{\reff}[1]{(\ref{#1})}
\newcommand{\bra}[1]{\left(#1\right)}
\newcommand{\bras}[1]{\left[#1\right]}
\newcommand{\brac}[1]{\left\{#1\right\}}
\newcommand{\sfr}[2]{{\textstyle\frac{#1}{#2}}}
\newcommand{\fr}[2]{{\frac{#1}{#2}}}

\begin{document}

\title[Charged multifluids]{Charged multifluids in general relativity}

\author{Mattias Marklund\dag, Peter K.\ S.\ Dunsby\ddag\S, Gerold
 Betschart\ddag, Martin Servin\P\ and Christos G.\ Tsagas\ddag}

\address{\dag\ Department of Electromagnetics, Chalmers University of
  Technology, SE--412 96 G\"oteborg, Sweden}
\address{\ddag\ Department of Mathematics and Applied Mathematics,
  University of Cape Town, 7701 Rondebosch, South Africa}
\address{\S\ South African Astronomical Observatory, Observatory
7925, Cape Town, South Africa}
\address{\P\ Department of Physics, Ume{\aa} University,
  SE--901 87 Ume{\aa}, Sweden}


\begin{abstract}
The exact 1+3 covariant dynamical fluid equations for a
multi-component plasma, together with Maxwell's equations are
presented in such a way as to make them suitable for a
gauge-invariant analysis of linear density and velocity
perturbations of the Friedmann-Robertson-Walker model. In the case
where the matter is described by a two component plasma where
thermal effects are neglected, a mode representing high-frequency
plasma oscillations is found in addition to the standard growing
and decaying gravitational instability picture. Further
applications of these equations are also discussed.
\end{abstract}



\section{Introduction}
Plasmas and electromagnetic fields have an established widespread
presence in the universe and are known to play an important role
in many astrophysical and cosmological processes. Although in most
cases plasma physics can be adequately addressed within the
Newtonian or the special relativistic framework, there are
occasions where general relativistic considerations should be
taken into account. The physics of the early universe offer a very
good example in this respect. General relativistic treatments
require the rigorous setup of a self-consistent set of equations to
describe the plasma dynamics. Moreover, when perturbative
techniques are employed, there are extra considerations, such as
those related to the gauge invariance of the approach.
In this paper, we will try to provide such a setup in the context
of cosmological fluid dynamics, leaving the possibility of a
kinetic-theory based description open for the future.\footnote{When
  analysing the CMB spectrum, the kinetic approach is used for the
  photons, while the electrons are treated as a fluid (their
  interaction is mediated via Thomson scattering). This is in contrast
to many Newtonian applications of plasma physics, where the particle
nature of the electromagnetic field is neglected, while electrons are
described using a kinetic treatment.}

A number of techniques can be used to analyse the equations
describing general-relativistic plasmas. Depending on the nature
of the problem one might employ analytical, numerical and/or
perturbative methods. Analytical results are usually based on
severe symmetry assumptions, which unavoidably restricts their
applicability. Moreover, the inherent nonlinearity of Einstein's
theory means that numerical techniques are also non-trivial to
apply. Thus, in many cases the most useful method is the
perturbative one, possibly combined with numerical methods.
In general, we may distinguish between two types of approach:

\textit{Non-gravitating plasmas on curved background spacetimes:}\
This method is probably best applied to astrophysical situations,
and effectively it comprises two sub-cases: (a) weak
gravitational fields, described by a single potential, or weak
gravitational waves; (b) strong gravitational fields, where one
uses exact solutions to Einstein's field equations for the
background. The ``membrane paradigm'' (see~\cite{Thorne}) is a
good example of a formalism which has been developed for this
purpose.

\textit{Self-gravitating plasmas:}\ In this case one takes into
account the plasma contribution to the total gravitational field.
This approach, which is more technically demanding than cases (a)
and (b) above, is applicable to early universe studies, when most
of the baryonic matter was ionised. Below, we will give some
examples of studies that have been based on the above described
techniques.

Considerable amount of work has been done on the interaction
between plasmas and gravity waves and on the use of
electromagnetic fields for the detection of gravitational waves
(see \cite{Braginski-etal,Demianski,Ignatev} and references
therein). This includes studies of parametric excitation of plasma
waves in the presence of gravitational
radiation~\cite{Brodin-Marklund}, the scattering of gravity waves
on highly energetic plasmas during supernovae
explosions~\cite{Bingham-etal}, and the possible existence of
radio waves due to the emission of weak gravitational waves from
binary pulsars~\cite{Marklund-Brodin-Dunsby}. Also, in analogy to
the frequency upshifting of short laser pulses observed in
laboratory plasmas (e.g.~see~\cite{Wilks-etal}), it was shown that
weak gravitational waves could induce similar phenomena in
magnetized multi-component plasmas~\cite{Brodin-Marklund-Servin}.
Moreover, in~\cite{Brodin-Marklund-Dunsby} the exact plane-fronted
parallel (pp) solution to Einstein's field equations
(e.g.~see~\cite{Kramer-Stephani-MacCallum-Herlt}) was employed to
gain a better understanding of nonlinearities in the interaction
between plasmas and gravitational waves (see
also~\cite{Ignatev2}).

A number of papers employ the membrane paradigm~\cite{Thorne},
together with the appropriate fluid equations, to look into the
plasma properties in the vicinity of compact astrophysical objects
such as black holes. In~\cite{Daniel-Tajima}, for example, the
authors studied high frequency EM-waves in plasma outside a
spherically symmetric black hole, and in~\cite{Daniel-Tajima2}
they show the possibility of an EM-wave outburst from black holes
due to mode conversion. Studies looking at the plasma behaviour
near rotating black holes can also be found in the
literature~\cite{Khanna}.

Work has also been done on fluid dynamics and kinetic gas theory
with the context of cosmology. Notably, the book by Bernstein
\cite{Bernstein}, which treats the gas kinetics in the
Friedmann-Lema\^itre-Robertson--Walker (FLRW) model. Nevertheless,
there are relatively few relativistic cosmological studies that
take into account plasma effects and the behaviour of matter in
the presence of electromagnetic
fields~\cite{PE,SB,TB1,Tsagas-Barrow,Tsagas-Maartens,TM2,Marklund-Dunsby-Brodin}.
Thus, the general relativistic treatment of plasmas, both in
astrophysics as well as in cosmology, looks like a field open to
investigation.

When studying relativistic cosmological perturbations, Bardeen's
gauge-invariant formalism is the most influential
approach~\cite{Bardeen}. However, Bardeen's theory is one of some
complexity and his variables do not always have a transparent
physical and geometrical interpretation. Moreover, the approach is
limited to linear perturbations around a FLRW background. Building
on the work of Hawking~\cite{H} and Olson~\cite{O} and utilising
that of Steward and Walker~\cite{Stewart-Walker}, Ellis and Bruni
\cite{Ellis-Bruni} introduced a mathematically simple and
physically transparent perturbation scheme. Their formalism, which
is both covariant and gauge-invariant, has the additional
advantage of not been confined to perturbed FLRW universes
(see~\cite{Ellis-vanElst} for a comprehensive review). The single
fluid analysis of Ellis and Bruni has been extended to
multi-component systems by Dunsby, Bruni and
Ellis~\cite{Dunsby-Bruni-Ellis}, where a number of possible
cosmological applications was discussed. Here, we will apply the
multi-component formalism of~\cite{Dunsby-Bruni-Ellis} to the case
of a charged two-component fluid.

\section{Preliminaries}
\subsection{The multi-component fluid}
We assume a family of fundamental observers moving with 4-velocity
$u^a$ and a collection of perfect fluids with individual
4-velocities given by
\begin{equation}
u_{(i)}^a = \gamma_{(i)}(u^a + v_{(i)}^a) \ ,
\end{equation}
where $\gamma_{(i)} \equiv (1 -v_{(i)}^2)^{-1/2}$ is the
Lorentz-boost factor and $v_{(i)}^au_a = 0$ ($i$ is numbering each
fluid). By assumption each fluid has, in its own rest frame, an
energy momentum tensor of the form
\begin{equation}\label{energymomentum1}
T_{(i)}^{ab} = (\mu_{(i)} + p_{(i)})u_{(i)}^au_{(i)}^b +
p_{(i)}g^{ab} \ ,
\end{equation}
where $\mu_{(i)}$ and $p_{(i)}$ are the fluid's energy density and
pressure respectively, while $g_{ab}$ is the spacetime metric.
Note that in general each species has its own equation of state.
Relative to the fundamental frame $u^a$, however, the above reads
\begin{equation}
T_{(i)}^{ab} = \hat{\mu}_{(i)} u^au^b + \hat{p}_{(i)}h^{ab} +
2u^{(a}\hat{q}_{(i)}^{b)} + \hat{\pi}_{(i)}^{ab} \ ,
\end{equation}
which is the stress-energy tensor of an imperfect fluid with
\numparts \label{energymomquantities}
\begin{eqnarray}
  \hat{\mu}_{(i)} \equiv
    \gamma_{(i)}^2(\mu_{(i)} + p_{(i)}) - p_{(i)} \ , \label{Density} \\
  \hat{p}_{(i)} \equiv
    p_{(i)} + \case{1}{3}\gamma_{(i)}^2(\mu_{(i)} + p_{(i)})v_{(i)}^2 \ ,
  \label{Pressure} \\
  \hat{q}_{(i)}^a \equiv
    \gamma_{(i)}^2(\mu_{(i)} + p_{(i)})v_{(i)}^a \ , \label{Heatflow} \\
  \hat{\pi}_{(i)}^{ab} \equiv
    \gamma_{(i)}^2(\mu_{(i)} + p_{(i)})(v_{(i)}^av_{(i)}^b -
    \case{1}{3}v_{(i)}^2h^{ab}) \ , \label{Aniso-pressure}
\end{eqnarray}
\endnumparts
and $h^{ab} \equiv g^{ab} + u^au^b$ is the projection tensor
orthogonal to $u^a$. Note that $\hat{q}_{(i)}^a$ is the heat flow
and $\hat{\pi}_{(i)}^{ab}$ is the anisotropic pressure of each
fluid component relative to $u^a$. Clearly, both quantities depend
entirely on the motion of the species relative to $u^a$.

\subsection{The electromagnetic field}
Charged fluids will interact with each other in the presence of an
electromagnetic field. Thus, we also assume the presence of an
electromagnetic field described by the Faraday tensor
\begin{equation}
F^{ab}=2u^{[a}E^{b]}+\epsilon^{abc}B_c\,,  \label{F}
\end{equation}
where $E^a=F^{ab}u_b$ and
$B^a={\textstyle{1\over2}}\epsilon^{abc}F_{bc}$ are respectively
the electric and magnetic fields as measured by the fundamental
observers ($\epsilon_{abc}$ is the spatial permutation tensor).
The electromagnetic field contributes to the total energy momentum
tensor by
\begin{equation}
\fl T_{\rm (em)}^{ab}=\case{1}{2}(E^2+B^2)u^au^b+
\case{1}{6}(E^2+B^2)h^{ab}+ 2u^{(a}\epsilon^{b)cd}E_cB_d-
(E^{\langle a}E^{b\rangle}+ B^{\langle a}B^{b\rangle})\,,
\end{equation}
where angled brackets indicate the projected, symmetric and
trace-free part of spacelike vectors and tensors. Finally, the
field obeys Maxwell's equations
\numparts \label{M}
\begin{eqnarray}
\nabla_bF^{ab}=\mu_0j^a\, , \label{M1}\\
\nabla_{[a}F_{bc]}=0\,.  \label{M2}
\end{eqnarray}
\endnumparts

\subsection{The gravitational field}
The dynamics of the gravitational field is determined by
Einstein's equations, forming a closed system once the equation of
state for the individual fluid components has been established. Of
course, in the presence of other physical fields (e.g.\
anisotropic stresses or spinor fields) we need to supplement the
system with the corresponding evolution and constraint equations
(e.g.~see~\cite{Tajima-Shibata,Marklund-Brodin-Shukla} and
references therein). In the presence of an electromagnetic field,
the conservation laws for the individual charged species are
\begin{equation}\label{Conservation}
\nabla_bT_{(i)}^{ab}=\frac{1}{\epsilon_0}F^a\!_b j_{(i)}^b+
J_{(i)}^a\,,
\end{equation}
with $j_{(i)}^a = \rho_{\mathrm{\rm c}(i)}u_{(i)}^a$ being the
4-current, $\rho_{\mathrm{\rm c}(i)} \equiv -u_aj_{(i)}^a$ the
charge density in the rest frame of the fluid and $\epsilon_0$ is
the permittivity of vacuum. The term $J_{(i)}^a$ represents
interactions other than electromagnetic between the fluids and
splits as
\begin{equation}
J_{(i)}^a = \varepsilon_{(i)}u^a + f_{(i)}^a \ ,
\end{equation}
where $\varepsilon_{(i)}$ is the work per unit volume due to the
interaction and $f_{(i)}^a$ is the force density orthogonal to
$u^a$. Because of overall energy-momentum conservation we require
that $\sum_i J_{(i)}^a=0$ and write the total fluid equations as
\begin{equation}\label{total}
\sum_i\nabla_bT_{(i)}^{ab} = \frac{1}{\epsilon_0}F^a\!_b \sum_i
j_{(i)}^b\,.
\end{equation}
Moreover, particle conservation ensures that
\begin{equation}\label{Massconservation}
\nabla_a(n_{(i)}u_{(i)}^a) = 0 \ ,
\end{equation}
where $n_{(i)}$ is the number density of the individual species in
their own rest frame. Finally, we point out that the current density
in Eq.\ (\ref{Conservation}) can be written $j_{(i)}^a =
q_{(i)}n_{(i)}u_{(i)}^a$, where $q_{(i)}$ is the individual charge
of the particles that make up the fluid.\footnote{In general, we
need to employ the second law of thermodynamics $\nabla_aS^a \ge
0$, supply an equation of state for the species and use the
covariant equations given in the Appendix.}

\section{The fluid equations}
\subsection{The nonlinear equations}
The conservation laws of the individual fluid components, relative
to the $u^a$ frame, are obtained by inserting decompositions
(\ref{Density})--(\ref{Aniso-pressure}) into
Eq.~(\ref{Conservation}). In particular, by projecting
(\ref{Conservation}) onto $u^a$ we arrive at the energy density
conservation equation
\begin{eqnarray}
\fl \dot{\mu}_{(i)}=-\left(\mu_{(i)}+p_{(i)}\right)
\left(\Theta+\tilde{\nabla}_av_{(i)}^a\right)-
\gamma_{(i)}^{-1}\left(\mu_{(i)}+p_{(i)}\right)
\left(\dot{\gamma}_{(i)}+\gamma_{(i)}\dot{u}_av_{(i)}^a
+v_{(i)}^a\tilde{\nabla}_a\gamma_{(i)}\right)\nonumber\\
-v_{(i)}^a\tilde{\nabla}_a\mu_{(i)}+
\gamma_{(i)}^{-1}\varepsilon_{(i)}\,. \label{Energy2}
\end{eqnarray}
On the other hand, we derive the momentum density conservation
equation
\begin{eqnarray}
\fl
\left(\mu_{(i)}+p_{(i)}\right)\left(\dot{u}^a+\dot{v}_{(i)}^{\langle
a\rangle}\right)=-\gamma_{(i)}^{-2}\tilde{\nabla}^ap_{(i)}-
\case{1}{3}\Theta\left(\mu_{(i)}+p_{(i)}\right)v_{(i)}^a-
\dot{p}_{(i)}v_{(i)}^a\nonumber\\ -\left(\mu_{(i)}+p_{(i)}\right)
\left(v_{(i)}^b\tilde{\nabla}_bv_{(i)}^a+\sigma^a\!_bv_{(i)}^b
+\epsilon^{abc}\omega_bv_{(i)c}\right)\nonumber\\
+\gamma_{(i)}^{-1}\left(\mu_{(i)}+p_{(i)}\right)
\left(v_{(i)}^a\dot{\gamma}_{(i)}
+v_{(i)}^av_{(i)}^b\tilde{\nabla}_b\gamma_{(i)}\right)\nonumber\\
-v_{(i)}^av_{(i)}^b\tilde{\nabla}_bp_{(i)}+
\gamma_{(i)}^{-1}\rho_{{\rm c}(i)}(E^a
+\epsilon^{abc}v_{(i)b}B_c)+ \gamma_{(i)}^{-1}f_{(i)}^a\,,
\label{Momentum2}
\end{eqnarray}
by projecting (\ref{Conservation}) orthogonal to $u^a$. Furthermore,
the particle number conservation, expressed by Eq.\
(\ref{Massconservation}), takes the form
\begin{equation}
\dot{n}_{(i)}=-\Theta n_{(i)}- n_{(i)}\dot{u}_av_{(i)}^a-
\gamma_{(i)}^{-1}\left[\dot{\gamma}_{(i)}n_{(i)}
+\tilde{\nabla}_a\left(\gamma_{(i)} n_{(i)}
v_{(i)}^a\right)\right]\,.  \label{Mass}
\end{equation}

Similarly, the total fluid equations (see Eq.~(\ref{total}))
provide the total energy density conservation,
\begin{equation}\label{totalenergy}
\dot{\mu}=-\Theta(\mu + p)-\tilde{\nabla}_aq^a- 2\dot{u}_aq^a-
\sigma^a\!_b\pi^b\!_a\,,
\end{equation}
and the total momentum density conservation
\begin{eqnarray}\label{totalmomentum}
\fl (\mu+p)\dot{u}^a=-\tilde{\nabla}^ap- \case{4}{3}\Theta q^a-
\dot{q}^{\langle a\rangle}- \sigma^a\!_bq^b-
\epsilon^{abc}\omega_bq_c- \tilde{\nabla}_b\pi^{ab}-
\dot{u}_b\pi^{ab}\nonumber\\ +\rho_{\rm c} E^a+
\epsilon^{abc}j_bB_c\,,
\end{eqnarray}
where $\rho_{\rm c}=\sum_i\rho_{{\rm c}(i)}$, $j^{\langle
b\rangle}=\sum_i j_{(i)}^{\langle b\rangle}$ are the total charge
and current density respectively. Also,
$\mu=\sum_i\hat{\mu}_{(i)}$, $p=\sum_i\hat{p}_{(i)}$, $q^a=\sum_i
\hat{q}_{(i)}^a$, $\pi^{ab}=\sum_i\hat{\pi}_{(i)}^{ab}$ by
definition and the hatted quantities are given by
(\ref{Density})--(\ref{Aniso-pressure}).

The covariant form of Maxwell's equations is obtained by
substituting the Faraday tensor, given by (\ref{F}), into
Eqs.~(\ref{M1}) and (\ref{M2}). They comprise a set of two
propagation and two constraint equations given
by~\cite{TB1,Tsagas-Barrow,Tsagas-Maartens,TM2}
\numparts\label{Maxwell}
\begin{eqnarray}
\dot{E}^{\langle a\rangle}&=&-\case{2}{3}\Theta E^a+
\sigma^a\!_bE^b+ \epsilon^{abc}\omega_bE_c+
\epsilon^{abc}\dot{u}_bB_c+ {\rm curl}B^a-
\frac{1}{\epsilon_0}j^{\langle a\rangle}\,, \label{Maxwell1}\\
\dot{B}^{\langle a\rangle}&=&-\case{2}{3}\Theta B^a+
\sigma^a\!_bB^b+ \epsilon^{abc}\omega_bB_c-
\epsilon^{abc}\dot{u}_bE_c-{\rm curl}E^a\,, \label{Maxwell2}\\
\tilde{\nabla}_aE^a&=&\frac{1}{\epsilon_0}\rho_{\rm c}+
2\omega_aB^a\,,  \label{Maxwell3}\\
\tilde{\nabla}_aB^a&=&-2\omega_aE^a\,,  \label{Maxwell4}
\end{eqnarray}
\endnumparts
where ${\rm curl}B^a\equiv\epsilon^{abc}\tilde{\nabla}_bB_c$, and
analogously for ${\rm curl}E^a$.

\subsection{The linear equations}
We will now linearize the equations of the previous section about
a FLRW model. Given the homogeneity and the isotropy of the FLRW
spacetime, all spatial gradients and velocity components
orthogonal to $u^a$ must vanish in the background. This implies
that spatial inhomogeneities are first order quantities and that
$\gamma_{(i)}=1$ to first order. In addition, the symmetries of
the FLRW background require that the electromagnetic field vanish
to zero order as well. This in turn implies, through Eq.\
(\ref{Maxwell3}), that $\rho_{\mathrm{c}}$ has zero background
value. Similarly, the shear $\sigma^{ab}$, vorticity $\omega^a$
and the acceleration $\dot{u}^a$ also vanish to zeroth order. As a
result, Eqs.\ (\ref{Energy2})--(\ref{Mass}) linearise as follows
\numparts
\begin{eqnarray}
\fl \dot{\mu}_{(i)}=-\left(\Theta+\tilde{\nabla}_av_{(i)}^a\right)
\left(\mu_{(i)}+p_{(i)}\right)\,,  \label{Energy3}\\
\fl
\left(\mu_{(i)}+p_{(i)}\right)\left(\dot{u}^a+\dot{v}_{(i)}^a\right)=
-\tilde{\nabla}^ap_{(i)}- v_{(i)}^a\dot{p}_{(i)}-
\case{1}{3}\Theta\left(\mu_{(i)}+p_{(i)}\right)v_{(i)}^a+
\rho_{{\rm c}(i)}E^a\,, \label{Momentum3}\\
\fl \dot{n}_{(i)}=
-\left(\Theta+\tilde{\nabla}_av_{(i)}^a\right)n_{(i)}\,,
\label{Mass2}
\end{eqnarray}
\endnumparts
where we have ignored non-electromagnetic interactions between the
species (i.e.~$\varepsilon_{(i)}=0=f_{(i)}^a$). Similarly, the
total fluid equations (\ref{totalenergy}) and
(\ref{totalmomentum}) reduce to
\numparts
\begin{eqnarray}
\dot{\mu}=-\Theta(\mu+p)- \tilde{\nabla}_aq^a&\,,  \label{totalenergy2}\\
(\mu + p)\dot{u}^a=-\tilde{\nabla}^ap- \case{4}{3}\Theta q^a-
\dot{q}^a\,.  \label{totalmomentum2}
\end{eqnarray}
\endnumparts
Finally, Maxwell's equations give \numparts\label{Maxwellpert}
\begin{eqnarray}
\dot{E}^a=- \case{2}{3}\Theta E^a+ {\rm curl}B^a-
\frac{1}{\epsilon_0}j^a\,, \label{Maxwell1b}\\
\dot{B}^a=-\case{2}{3}\Theta B^a- {\rm curl}E^a\,,
\label{Maxwell2b}\\ \tilde{\nabla}_aE^a=
\frac{1}{\epsilon_0}\rho_{\rm c}\,,  \label{Maxwell3b}\\
\tilde{\nabla}_aB^a = 0\,.  \label{Maxwell4b}
\end{eqnarray}
\endnumparts

In many applications, it has proved advantageous to adopt the
energy frame, defined by the vanishing of the energy
flux,\footnote{The electromagnetic contribution to the total heat
flux through the Poynting vector $\epsilon^{abc}E_bB_c$ vanishes
to first order.}
\begin{equation}
q^a = \sum_i \hat{q}_{(i)}^a = 0 \ .
\end{equation}
In this frame Eq.\ (\ref{totalmomentum}) reduces to
\begin{equation}
(\mu + p)\dot{u}^a=-\tilde{\nabla}^ap\,,
\end{equation}
which means that for a dust background the acceleration vanishes
to first order.

\section{Applications}
\subsection{Electrically induced velocity perturbations}
\label{examples}
Consider an Einstein-de Sitter background and a two-fluid system,
with each component having a dust-like energy-momentum tensor
relative to its own frame. In the background, the only non-zero
scalars are the total density $\mu = \mu_1 + \mu_2$ and the
expansion $\Theta$. Note that to zero order the total charge
vanishes (i.e.~$\rho_{\rm c}=-e(n_1-n_2)=0$), since both species
have equal but opposite charges $q_1=-e=-q_2$. It follows that
$\rho_{\rm c}$ is a first order gauge-invariant
variable~\cite{Stewart-Walker}. Furthermore, $\mu_i=m_in_i$ since
no thermal effects are included. In this environment, it is useful to
introduce the variables
\begin{equation}
N=n_1 + n_2\,, \hspace{3mm} n=n_1-n_2\,, \hspace{3mm}
V^a=\case{1}{2}(v^a_1+v^a_2)\,, \hspace{3mm}
v^a=\case{1}{2}(v^a_1-v^a_2)\,.
\end{equation}
\endnumparts
Given our frame choice (i.e.~$q^a=0$), Eq.\ (\ref{Heatflow})
leads to the first order result $\mu_1 v_1^a = -\mu_2 v_2^a$ and
subsequently to the following relation
\begin{equation}
  V^a = -\frac{\delta\mu}{\mu}v^a \ ,
\end{equation}
between $V^a$ and $v^a$, where $\delta\mu = \mu_1 - \mu_2$ and
$\delta\mu/\mu$ is independent of time. Then, employing Eqs.\
(\ref{Momentum3}) and (\ref{Mass2}) we obtain the propagation
formulae for $N$, $n$ and $v^a$
\numparts
\begin{eqnarray}
\dot{N}&=&-\left(\Theta+\tilde{\nabla}_aV^a\right)N\,,
\label{pert1}\\
\dot{n}&=&-\Theta n- N\tilde{\nabla}_av^a\,, \label{pert3}\\
\dot{v}^{a}&=&-\case{1}{3}\Theta v^a-
\frac{e}{2}\frac{(m_1+m_2)}{m_1m_2}E^a\,. \label{pert2}
\end{eqnarray}
\endnumparts
As expected, Eqs.~(\ref{pert1}) and (\ref{pert3}) show how velocity
perturbations, depending on the sign of their 3-divergence, can
increase or decrease the number density dilution caused by the
expansion. More importantly, Eq.~(\ref{pert2}) shows that the
presence of the electric field acts as a source of linear velocity
perturbations in the charged plasma, even when such perturbations
are originally absent (i.e.~when $v_a=0$ initially). In what follows
we will see that a non-zero initial velocity perturbation can give rise
to density fluctuations (cf.~\reff{pert3}), which through Eq.~\reff{Maxwell3b}
may seed electric fields.

\subsection{Velocity induced density perturbations}

Consider the dimensionless, first-order, gauge-invariant variable
\begin{equation}
\Delta=\frac{a^2}{N}\tilde{\nabla}^2N\,,  \label{Delta}
\end{equation}
where $a$ is the background scale factor and
$\tilde{\nabla}^2=h^{ab}\3nab_a\3nab_b$ is the covariant
Laplacian operator normal to $u^a$. The above describes
inhomogeneities in the total number density of the particles and,
consequently, it also describes inhomogeneities in the total
energy density. To linear order the evolution of $\Delta$ is
determined by the system
\numparts
\begin{eqnarray}
\label{density1}
  \dot{\Delta} &=& -{\mathscr{Z}}
    + \frac{\delta\mu}{\mu}a\tilde{\nabla}^2{\mathscr{V}}\ , \\
\label{expansion}
  \dot{{\mathscr{Z}}} &=&
    -\case{2}{3}\Theta{\mathscr{Z}}
    - \case{1}{4}N\left[ (m_1 + m_2)\Delta
      + (m_1 - m_2)a^2\tilde{\nabla}^2Y\right] \ ,\\
\label{density2}
  \dot{{\mathscr{V}}} &=&
    -\case{1}{3}\Theta{\mathscr{V}}
      + \case{3}{4}\alpha^2\mu aY\ , \\
\label{numberden} \dot{Y}&=&-\frac{1}{a}{\mathscr{V}}\ ,
\end{eqnarray}
\endnumparts
where $\alpha^2=4e^2/3\epsilon_0m_1m_2$. In deriving the above we
have employed the first order gauge-invariant variables
\begin{equation} \label{defs}
{\mathscr{Z}}=a^2\tilde{\nabla}^2\Theta\,, \hspace{5mm}
{\mathscr{V}}=a\tilde{\nabla}_av^a\,, \hspace{5mm} Y=n/N\,,
\end{equation}
and used Maxwell's equation (\ref{Maxwell3}). Note that
${\mathscr{Z}}$ and ${\mathscr{V}}$ describe scalar
inhomogeneities in the expansion and the relative velocity of the
species respectively, while $Y$ determines the net charge of the
total fluid. Given that Eqs.~(\ref{density2}) and (\ref{numberden})
have decoupled from the rest of the system we can
obtain the following propagation equation for $Y$:
\begin{equation}
  \ddot{Y}
  + \case{2}{3}\Theta\dot{Y}
  + \case{3}{4}\alpha^2\mu Y= 0\;,
\label{Yequation}
\end{equation}
The solution to Eq.\ (\ref{Yequation}) will act as an
inhomogeneous driving term in the corresponding propagation equation
for $\Delta$:
\begin{equation}
  \ddot{\Delta}  + \case{2}{3}\Theta\dot{\Delta}-\case{1}{2}\mu\Delta
=\left(\case{3}{4}\alpha^2+\case{1}{2}\right)\frac{\delta\mu}{\mu}\mu
a^2\tilde{\nabla}^2Y\ ,
\label{Dequation}
\end{equation}
obtained by taking the derivative of Eq.\ (\ref{density1}) and
using (\ref{expansion}). According to Eqs.~(\ref{numberden}) and
(\ref{Dequation}),
 velocity
inhomogeneities act as sources of density fluctuations. Note that
the right hand side of \reff{Dequation} is a pure multifluid
effect, where the part containing $\alpha^2$ stems from the plasma
description.

In order to solve equations (\ref{Yequation}) and
(\ref{Dequation}) it is standard to decompose the physical
(perturbed) fields into a spatial and temporal part, using as
eigenfunctions $Q_{(k)}$, solutions of the scalar Helmholtz
equation \cite{bi:harrison}. In particular we write
\begin{equation}
\Delta=\Delta_{(k)}Q^{(k)}\,, \hspace{10mm} Y=Y_{(k)}Q^{(k)}\,,
\end{equation}
where $\tilde{\nabla}_aY_{(k)}=0=\tilde{\nabla}_a\Delta_{(k)}$,
$\dot{Q}_{(k)}=0$ and $\tilde{\nabla}^2Q^{(k)}=-(k^2/a^2)Q^{(k)}$.
For an Einstein-de Sitter background, the expansion and energy
density evolve as $\Theta=2/t$ and $\mu=4/3t^2$. Hence, applying
the harmonic splitting given above, Eqs.~(\ref{Dequation}) and
(\ref{Yequation}) become
\begin{equation}
\ddot{\Delta}_{(k)}
  + \frac{4}{3t}\dot{\Delta}_{(k)}
  - \frac{2}{3t^2}\Delta_{(k)} = -\case{1}{3}k^2(3\alpha^2+2)
   \frac{\delta\mu}{\mu}\frac{1}{t^2}Y_{(k)}\,,
\end{equation}
and
\begin{equation} \label{densitywave2}
 \ddot{Y}_{(k)} + \frac{4}{3t}\dot{Y}_{(k)}
  +\frac{\alpha^2}{t^2}Y_{(k)} = 0 \ ,
\end{equation}
respectively. In order to estimate the value of the parameter
$\alpha$ we substitute back for the gravitational constant and
write
\begin{equation}
\alpha^2=\frac{4}{3}\left(\frac{m_e}{m_1}\right)\left(\frac{m_e}{m_2}\right)
\left(\frac{e^2}{\epsilon_0}\right)\left(\frac{1}{8\pi
Gm_e^2}\right)\sim\left(\frac{m_e}{m_1}\right)\left(\frac{m_e}{m_2}\right)
\times 10^{42}\;.
\end{equation}
Since $\alpha\gg 1$ the solutions to above equations are
\numparts
\bea
\Delta_{(k)}&=&{\cal C}_1\tau^{2/3}+ {\cal C}_2\tau^{-1}+
k^2\frac{\delta\mu}{\mu}Y_{(k)}\;,\\
Y_{(k)}&=&\left[{\rm C}_1\cos(\alpha\ln \tau)+{\rm
C}_2\sin(\alpha\ln \tau)\right]\tau^{-\case{1}{6}}\;,
\eea
\endnumparts
where we have introduced the dimensionless time-coordinate $ \tau
\equiv t/t_i$, with $t_i$ corresponding to some arbitrary initial
time. Hence, in addition to the usual growing and decaying modes
of the standard gravitational instability picture, we have
obtained a mode representing high frequency plasma
oscillations with a weak damping envelope. This mode is triggered
by velocity distortions in the charged plasma and, as expected,
has negligible large scale effect. However, the extra plasma modes
become increasingly important as we move on to progressively
smaller scales (i.e.~for $k\gg1$).

It should also be pointed out that a \emph{finite temperature} will in
general cause Landau damping of the plasma oscillations. The effect
(requiring kinetic treatment) is small for wavelengths much larger than
the Debye length (which is proportional to the thermal velocity of the
plasma particles) and in this case the dust fluid approximation is well
justified.


\subsection{Velocity induced electromagnetic fields}

In this section, we will derive the wave equations for the
electromagnetic field, seeded by velocity perturbations.

For a cold plasma, the currents for each fluid species may be
written as
\be
j_{(i)}^a = q_{(i)}n_{(i)}u_{(i)}^a = q_{(i)}n_{(i)}(u^a+
v_{(i)}^a)\ , \label{currents}
\ee
where $q_{(i)}$ is the charge and $v_{(i)}^a$ is the velocity of
the species under consideration. Since we require the plasma to be
neutral on the whole, the species are of opposite charge. Hence,
the total current $j^a$ appearing in Maxwell's equations reads to
first order

\be
j^a = j_1^a + j_2^a = -e N v^a\ . \label{totalcurrent}
\ee
>From Maxwell's equations \reff{Maxwell1b}--\reff{Maxwell4b}, using \reff{totalcurrent} and \reff{pert2}, one
can then deduce wave equations for the induced
electromagnetic fields:
\numparts
\bea  \fl
 \ddot{E}_{\la a\ra} - \3nab^2 E_a +\sfr{5}{3}\Theta\dot{E}_{\la a\ra}
 +\bras{\sfr{2}{9}\Theta^2
 +\bra{\sfr{3}{4}\alpha^2+\sfr{1}{3}}\mu}E_a
 =2\beta^2\mu\bra{\3nab_a Y-\sfr{1}{3}\Theta
 v_a}\ ,\label{waveE} \\ \fl
  \ddot{B}_{\la a\ra} - \3nab^2 B_a +\sfr{5}{3}\Theta\dot{B}_{\la a\ra}
 +\bras{\sfr{2}{9}\Theta^2 +\sfr{1}{3}\mu}B_a
 = -2\beta^2\mu \,\c v_a\ , \label{waveB}
\eea
\endnumparts
where $\beta^2 \equiv e/\epsilon_{0}(m_1 + m_2)$.
These equations govern the interaction of density/velocity
perturbations with electromagnetic waves in the plasma and show in
particular that density/velocity perturbations induce wave phenomena.
Observe that $B_a$ and $\c\,v_a$ are both purely solenoidal,
whereas $\3nab_a Y$ has no solenoidal  part. It is worthwhile to
note that the magnetic field is solely sourced by inhomogeneities
in the velocity in contrast to the electric field which is sourced
by inhomogeneities in the number density and velocity
perturbations. Clearly, the velocity perturbation is non-zero even
if $E^a = 0$, as long as $v^a \neq 0$ initially (cf.\,\reff{pert2}).
Both Eqs.~(\ref{waveE}) and \reff{waveB}  look strikingly similar, the
differences originating either from the total current or from a
gradient in the charge density (in the case of $\3nab_a Y$). The
additional $3\alpha^2/4$-term in the electric wave equation comes
from the non-stationarity of the total current, and its large magnitude ---
$\alpha^2 \sim 10^{42}$ for an $e^+e^-$-plasma --- leads directly
to the high-frequency behaviour of plasma effects, as will be
shown below (see also the preceding section).

It will be useful to introduce expansion normalized variables,
\be \mathscr{E}_a \equiv \fr{E_a}{\Theta}\ ,\;\;\;\mathscr{B}_a \equiv
\fr{B_a}{\Theta}\ ,\;\;\;\mathscr{K}_a \equiv \fr{\c
v_a}{\Theta}\ .\label{exp} \ee
Equations \reff{waveE} and \reff{waveB}, together with equations
for the driving terms, then read
\numparts
\bea
\fl  \ddot{\mathscr{E}}_{\la a\ra} - \3nab^2 \mathscr{E}_a  +\bra{\Theta-\sfr{\mu}{\Theta}}
  \dot{\mathscr{E}}_{\la a\ra}  +\bras{-\sfr{1}{9}\Theta^2
 +\bra{\sfr{3}{4}\alpha^2+\sfr{1}{3}}\mu}\mathscr{E}_a= 2\beta^2 \sfr{\mu}{\Theta}(\3nab_a Y-\sfr{1}{3}\Theta v_a)\ ,
\label{wavecalE} \\ \fl
  \dot{v}_{\la a \ra} + \sfr{1}{3}\Theta v_a = -\sfr{3}{8} \sfr{\alpha^2}{\beta^2} \Theta \mathscr{E}_a    \label{vel}   \ ,\\ \fl
  \ddot{\mathscr{B}}_{\la a\ra} - \3nab^2 \mathscr{B}_a +\bra{\Theta -\sfr{\mu}{\Theta}}\dot{\mathscr{B}}_{\la a\ra}
 +\bras{-\sfr{1}{9}\Theta^2 +\sfr{1}{3}\mu}\mathscr{B}_a= -2\beta^2\mu \,\mathscr{K}_a \;,
 \label{wavecalB}\\ \fl
 \dot\mathscr{K}_{\la a \ra}  +
 \bra{\sfr{1}{3}\Theta-\sfr{1}{2}\sfr{\mu}{\Theta}}\mathscr{K}_a
 = \sfr{3}{8}\sfr{\alpha^2}{\beta^2}\bras{\dot\mathscr{B}_{\la a \ra}+
 \bra{\sfr{1}{3}\Theta-\sfr{1}{2}\sfr{\mu}{\Theta}}\mathscr{B}_a}\ .
 \label{Kdrive}
\eea
\endnumparts
Equation \reff{Kdrive} follows from \reff{exp} using \reff{vel}
and Maxwell's equation \reff{Maxwell2b}.

Restricting ourselves to
{\it scalar} perturbations, we take the divergence of the above equations to extract
the scalar part of the system. Of course, there is no contribution
from the magnetic field in this case. Using \reff{defs} and defining $\mathscr{E} \equiv a\3nab^a\mathscr{E}_a$,
Eq.~\reff{vel} then transforms
into (cf.~\reff{density2})
\be \dot{\mathscr V} + \sfr{1}{3}\Theta \mathscr V = -\sfr{3}{8}
\sfr{\alpha^2}{\beta^2} \Theta \mathscr{E} = \sfr{3}{4}\alpha^2\mu a
Y\ , \label{vels} \ee
where the last equality is a direct consequence of Maxwell's
equation \reff{Maxwell3b}. Combining Eq.~\reff{numberden} with
\reff{vels} and using \reff{totalenergy2} together
with the commutator expression
\be
a\3nab^a\3nab^2\mathscr{E}_a = \3nab^2\mathscr{E}
+\bra{-\sfr{2}{9}\Theta^2 +\sfr{2}{3}\mu}\mathscr{E}\ ,
\ee
one can show that
the scalar part of the electric wave Eq.~\reff{wavecalE}
reduces to\footnote{Note that in deriving Eq.~\reff{wavecalEs}, the
  Laplacian terms cancel, and a harmonic decomposition is therefore
  not needed. Thus, the electric field will not contain a particular
  length scale, due to its Coulomb-like nature.}
\be
\ddot{\mathscr{E}} +\bra{\sfr{4}{3}\Theta-\sfr{\mu}{\Theta}}
\dot{\mathscr{E}} +\bras{\sfr{2}{9}\Theta^2
 +\bra{\sfr{3}{4}\alpha^2 -\sfr{1}{2}}\mu}\mathscr{E}
 = 0\ .\label{wavecalEs}
 \ee
In addition, Eq.~\reff{vels} gives rise to propagation
equations for $\mathscr V$ and $Y$, as discussed earlier:
\numparts
\bea
\ddot{\mathscr V}+\sfr{1}{3}\Theta\dot{\mathscr V}+\bras{-\sfr{1}{9}\Theta^2
+\bra{\sfr{3}{4}\alpha^2-\sfr{1}{6}}\mu}\mathscr V =0\ , \label{vprop} \\
\ddot{Y}+\sfr{2}{3}\Theta \dot{Y} +\sfr{3}{4}\alpha^2\mu Y=0\ .
\label{Yagain}
\eea
\endnumparts
Hence, Eqs.~\reff{wavecalEs}-\reff{Yagain} all stem from
\reff{vels}.

Specialising to a flat FLRW model with a zero cosmological constant, for which $\mu=1/3\,\Theta^2$ and $\Theta=2/t$
always holds, solutions to these equations can easily be obtained:
\numparts
\bea \fl \mathscr{V}(\tau) = \sfr{1}{\sqrt{\tau}}\brac{A \cos(\omega
\ln\tau)+\sfr{1}{\omega}\bra{\sfr{1}{2}A+B} \sin(\omega \ln\tau)}\ ,
\label{solV}\\ \fl \mathscr{E}(\tau) =
-\sfr{9}{4}\sfr{\beta^2}{\alpha^2}\sfr{1}{\sqrt{\tau}}\brac{\bra{2A+3B}\cos(\omega
\ln\tau)+\sfr{(2-18\alpha^2)A+3B}{6\omega}\sin(\omega \ln\tau)}\ ,
\label{solE} \\ \fl Y(\tau)
=\sfr{t_i}{3a_i}\sfr{1}{\alpha^2}\sfr{1}{\tau^{1/6}}\brac{\bra{2A+3B}\cos(\omega
\ln\tau)+\sfr{(2-18\alpha^2)A+3B}{6\omega}\sin(\omega \ln\tau)}\ .
\label{solY} \eea
\endnumparts
Here, we used again the dimensionless time-coordinate $ \tau
\equiv  t/t_i $, where $t_i$ denotes some arbitrary initial time.
Initial conditions of the velocity perturbation are chosen to be $A=\mathscr V(1)$
and $B=\mathscr V^{\prime}(1)$ (a prime stands for $\partial_\tau$). The
frequency of the solutions is proportional to $\omega \equiv
\sqrt{\alpha^2-1/36}$ and grows logarithmically in time. The solutions
exhibit high-frequency plasma behaviour. Observe that
although the solutions decay with time, their magnitude changes only
very slowly over time, particularly if the velocity perturbations are
taken to start at the onset of recombination.

We have restricted our attention to scalar perturbations, with the
implication that magnetic field effects vanish. From the point of view
of generating magnetic seed fields, for, e.g., the dynamo mechanism or
the Biermann-battery effect (see \cite{Widrow} and references
therein), it is of interest to analyse vector perturbations in a
similar way. This is reserved for future research.

\section{Discussion}
In this paper we generalized the multi-component fluid equations
derived by Dunsby, Bruni \& Ellis \cite{Dunsby-Bruni-Ellis} to the
case of charged fluids in the presence of electromagnetic fields. The
equations are given in covariant form, relative to an observer
moving with velocity $u^a$ that is taken to coincide with the
average velocity of the cosmic medium. We linearized these
equations about a FRW universe and then applied them to an
Einstein-de Sitter (EdS) universe. Our matter field is an
ion-electron plasma with zero average pressure (which made the EdS
model a suitable background). We showed how, when there is a
residual net charge, the presence of an electric field can lead to
velocity perturbations even when the latter are originally absent.
We also found that velocity distortions can source inhomogeneities
in the number density, and therefore in the energy density, of the
fluid. In fact, our linear equations reveal the presence of an
extra mode, representing high frequency plasma oscillations, in
addition to the standard growing and decaying modes. This mode is
likely to be important on scales considerably smaller than the
Hubble radius and therefore is of little importance as far as
structure formation is concerned. It does illustrate, however,
interesting small scale physics that could play a role during the
latter stages of galaxy formation.


We also applied our covariant equations to look into the
generation of electromagnetic fields due to velocity perturbations
in a plasma. The corresponding wave equations, with the velocity
distortions playing the role of a source, were given, and  they were solved in
the case of scalar perturbations. The solutions show high-frequency
behaviour typical of a plasma. We restricted our attention to scalar
perturbations, thus obtaining an evolution equation for the electric
field. However, magnetic field effects were absent since these are
related to vector modes. Because magnetic seed fields play a crucial
role in, e.g., the dynamo mechanism, it is of great interest to pursue
the analysis of the presented equations in the context of vector
perturbations. Results in this direction will be presented elsewhere.

There are a number of ways to generalize the discussion presented
in this paper. One possibility is to include thermal effects which
occur in a photon-baryon plasma giving a non-zero acceleration to
first order. This may lead to possible coupling between acoustic
and plasma oscillations. In addition one could apply the
ponderomotive force concept between neutrinos and electrons (see
\cite{Tajima-Shibata,Silva-etal} and references therein) to
cosmology in a covariant context. In this picture, derived from
the theory of electroweak interactions, there is an effective
interaction between electrons and neutrinos due to density
gradients in either species. For instance, the (non-relativistic)
force density exerted by neutrinos on the electrons is given by
\cite{Silva-etal}
\begin{equation} \label{neutrino}
  f_{(e)}^a = -\frac{1}{\sqrt{2}}\left( 1 + 4\sin^2\theta_W\right)%
    G_Fn_e\tilde{\nabla}^a n_{\nu} \ ,
\end{equation}
where $\theta_W$ is the Weinberg angle and $G_F$ is the Fermi
constant. The expression (\ref{neutrino}), together with its neutrino
counterpart, could act as a driving force for density fluctuations in
the early Universe, possibly giving a neutrino signature in the CMB,
having an alternating structure as compared to the regular CMB
spectrum.
The neutrino-driven instability discussed by Silva et al.\
\cite{Silva-etal2} (see also Ref.\ \cite{Semikoz} for the covariant
relativistic form of the same equations), using kinetic theory,
could in principle be transferred to a gauge invariant covariant
formalism, suitable for cosmological applications (see also
\cite{Misner}), but this is left for future studies.

\ack
This work was supported by Sida/NRF. M.M.\ would like to thank the
Cosmology Group at the Department of Mathematics and Applied
Mathematics, University of Cape Town, for their hospitality.
%
%

\appendix

\section{Gravitational dynamics}

The covariant equations for the dynamics of the gravitational field
was given in Ref.\ \cite{Ellis-vanElst}, and we use their notation.

\subsection{Covariant equation}

$\bullet$ Evolution equations for kinematic variables:
\begin{eqnarray}
\label{eq:ray}
\fl \dot{\Theta} - \3nab_{a}\udot^{a}
= - \,\case{1}{3}\,\Theta^{2} + (\udot_{a}\udot^{a})
- 2\,\sigma^{2} + 2\,\omega^{2} - \case{1}{2}\,(\mu+3p)
+ \Lambda \ , \\
\label{eq:omdot}
\fl \dot{\omega}^{\lgl a\rgl} - \case{1}{2}\,\eta^{abc}\,
\3nab_{b}\udot_{c}
= - \,\case{2}{3}\,\Theta\,\omega^{a} + \sigma^{a}\!_{b}\,\omega^{b} \ , \\
\label{eq:sigdot}
\fl \dot{\sigma}^{\lgl ab\rgl} - \3nab{}^{\lgl a}\udot^{b\rgl}
= - \,\case{2}{3}\,\Theta\,\sigma^{ab} + \udot^{\lgl a}\,
\udot^{b\rgl} - \sigma^{\lgl a}\!_{c}\,\sigma^{b\rgl c}
- \omega^{\lgl a}\,\omega^{b\rgl} - (E^{ab}-\case{1}{2}\,\pi^{ab}) \ ,
\end{eqnarray}

\noindent $\bullet$ Constraint equations for kinematic variables:
\begin{eqnarray}
0 = \3nab_{b}\sigma^{ab} - \case{2}{3}\,\3nab^{a}\Theta
+ \eta^{abc}\,[\ \3nab_{b}\omega_{c} + 2\,\udot_{b}\,\omega_{c}\ ]
+ q^{a} \ , \\
0 = \3nab_{a}\omega^{a} - (\udot_{a}\omega^{a}) \ , \\
\label{hconstr}
0 = H^{ab} + 2\,\udot^{\lgl a}\,
\omega^{b\rgl} + \3nab^{\lgl a}\omega^{b\rgl}
- (\c\,\sigma)^{ab} \ ,
\end{eqnarray}
where
$(\c\,\sigma)^{ab} = \eta^{cd\lgl a}\,\3nab_{c}\sigma^{b\rgl}\!_{d}$.

\noindent $\bullet$ Evolution equations for the curvature variables:
\begin{eqnarray}
\fl (\dot{E}^{\lgl ab\rgl}+\case{1}{2}\,\dot{\pi}^{\lgl ab\rgl})
- (\c\,H)^{ab} + \case{1}{2}\,\3nab^{\lgl a}q^{b\rgl}
 =  - \,\case{1}{2}\,(\mu+p)\,\sigma^{ab}
- \Theta\,(E^{ab}+\case{1}{6}\,\pi^{ab}) \nonumber \\
\fl  + \ 3\,\sigma^{\lgl a}\!_{c}\,(E^{b\rgl c}
-\case{1}{6}\,\pi^{b\rgl c}) - \udot^{\lgl a}\,q^{b\rgl}
+ \ \eta^{cd\lgl a}\,[\ 2\,\udot_{c}\,H^{b\rgl}\!_{d}
+ \omega_{c}\,(E^{b\rgl}\!_{d}+\case{1}{2}\,\pi^{b\rgl}\!_{d})\ ] \ ,
\\
\fl \dot{H}^{\lgl ab\rgl} + (\c\,E)^{ab} -\case12(\c\,\pi)^{ab}
 - \,\Theta\,H^{ab} + 3\,\sigma^{\lgl a}\!_{c}\,H^{b\rgl c}
+ \case{3}{2}\, \omega^{\lgl a}\,q^{b\rgl} \nonumber \\
- \ \eta^{cd\lgl a}\,[\ 2\,\udot_{c}\,E^{b\rgl}\!_{d}
- \case{1}{2}\,\sigma^{b\rgl}\!_{c}\,q_{d}
- \omega_{c}\,H^{b\rgl}\!_{d}\ ] \ ,
\end{eqnarray}
where
\begin{eqnarray}
(\c\,H)^{ab} =  \eta^{cd\lgl a}\,\3nab_{c}H^{b\rgl}\!_{d} \ , \\
(\c\,E)^{ab} = \eta^{cd\lgl a}\,\3nab_{c}E^{b\rgl}\!_{d} \ , \\
(\c\,\pi)^{ab} =  \eta^{cd\lgl a}\,\3nab_{c}\pi^{b\rgl}\!_{d} \ .
\end{eqnarray}

\noindent $\bullet$ Constraint equations for the curvature variables:
\begin{eqnarray}
\label{eq:divE}
\fl 0 =  \3nab_{b}(E^{ab}+\case{1}{2}\,\pi^{ab})
- \case{1}{3}\,\3nab^{a}\mu + \case{1}{3}\,\Theta\,q^{a}
- \case{1}{2}\,\sigma^{a}\!_{b}\,q^{b} - 3\,\omega_{b}\,H^{ab}
\nonumber \\
- \ \eta^{abc}\,[\ \sigma_{bd}\,H^{d}\!_{c}
- \case{3}{2}\,\omega_{b}\,q_{c}\ ] \ , \\
\label{eq:divH}
\fl 0 =  \3nab_{b}H^{ab} + (\mu+p)\,\omega^{a}
+ 3\,\omega_{b}\,(E^{ab}-\case{1}{6}\,\pi^{ab}) \nonumber \\
+ \ \eta^{abc}\,[\ \case{1}{2}\,\3nab_{b}q_{c}
+ \sigma_{bd}\,(E^{d}\!_{c} +\case{1}{2}\,\pi^{d}\!_{c})\ ] \ .
\end{eqnarray}

\section*{References}


\end{document}